\documentclass[english,aps,prl,twocolumn,showpacs,superscriptaddress,groupedaddress,amssymb]{revtex4-1}

\usepackage{graphicx}

\include{formats}

\begin{document}

\title{Electroactuation with Single Charge Carrier Ionomers}

\author{Alpha A. Lee }
\affiliation{Department of Chemistry, Faculty of Natural Sciences, Imperial College
London, SW7 2AZ, UK}
\author{Ralph H. Colby }
\affiliation{Department of Chemistry, Faculty of Natural Sciences, Imperial College
London, SW7 2AZ, UK}
\affiliation{Department of Materials Science and Engineering and Materials Research
Institute, The Pennsylvania State University, University Park, PA
16802, USA }

\author{Alexei A. Kornyshev}
\email{a.kornyshev@imperial.ac.uk}
\affiliation{Department of Chemistry, Faculty of Natural Sciences, Imperial College
London, SW7 2AZ, UK}

\begin{abstract}
A simple theory of electromechanical transduction for single-charge-carrier double-layer electroactuators is developed, in which the ion distribution and curvature are mutually coupled. The obtained expressions for the dependence of curvature and charge accumulation on the applied voltage, as well as the electroactuation dynamics, are compared with literature data. The mechanical- or sensor- performance of such electroactuators appears to be determined by just three cumulative parameters, with all of their constituents measurable, permitting a scaling approach to their design.\end{abstract}

\maketitle

Electro-mechanical materials have long intrigued physicists. Such materials can serve as sensors and transducers (where mechanical stimulation generates a voltage) or can function as electroactuators where voltage generates a mechanical response (for review see \cite{Oguro1992,Shahinpoor:1998jt,Bar-Cohen:2008hb}).
A class of such materials, called Ionomeric Polymer Metal Composites (IPMC), is based on the motion of ions of one
sign in an elastic polymer membrane enclosed between two bendable electrodes. Applied voltage causes mobile
ions to accumulate near the oppositely charged electrode, forming an electrical double layer that expands the membrane and a depletion layer at the opposite electrode that contracts the membrane, causing bending (Fig. 1). 
\begin{figure}[h]
\includegraphics[scale=0.4]{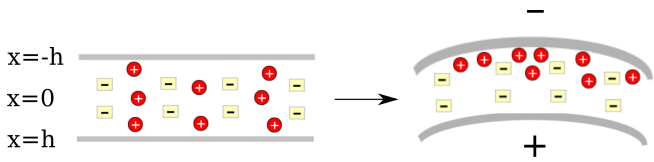}
\caption{Electroactuators based on mobile cations: redistribution of cations in the ionomer membrane between two flexible electrodes causes bending \cite{Bar-Cohen:2008hb}. }
\end{figure}

Early materials were swollen with polar solvents such as water, and were often single-ion conductors \cite{Oguro1998,Akle:2005gh,Leo2007}.
More recently, these volatile solvents were replaced with ionic liquids
\cite{Bennett:2004gf,Wang:2007ve,Duncan:2008ul,Liu2010} to eliminate evaporation problems, but necessarily creating dual-ion conductors that often bend first forward and later backwards under a constant applied voltage \cite{Lin2011}.
Sophisticated models for ionic actuation with multiple sorts of charge carriers have been developed \cite{Nemat-Nasser:2002zr,NematNasser2006b,FarinholtLeo2008,Pugal2010}. Those pioneering works, however, did not incorporate the feedback of mechanical deformation on the ionic charge distribution, which impedes their application for a description of the sensing function of electroactuators. The complexity of those models make them impractical for direct use in designing new actuators.

We focus here on a simpler system, containing only one mobile charge carrier \footnote{Single-charge-carrier ionomers are usually made of polyelectrolyte membrane with side-chains terminated by acid groups which dissociate leaving anionic groups bound to the side chains. The mobile protons can be chemically exchanged for various cations.} (for optimal electroactuation) for which it was straightforward to incorporate the stress-to-voltage feedback, as well as the volume of ions in electrical double layers. The analysis shows which physical characteristics determine the response and how they can be grouped into only two dimensionless cumulative parameters to describe the beam deflection under the applied voltage, with a third cumulative parameter determining the time scale of bending dynamics. This grouping reveals the interrelations between the physical characteristics, instructs what should be measured to predict electroactuator performance, and makes possible a scaling approach to their design. 
 
\emph{Statics} - The equilibrium properties of this system can be derived by minimising the free energy functional 
\begin{eqnarray}
\mathcal{F}[\psi,\rho,\kappa] & = & \mathcal{S}\int_{-h}^{h}\mathrm{dx}\left[-\frac{\varepsilon}{8 \pi}\left(\frac{\mathrm{d}\psi}{\mathrm{d}\mathrm{x}}\right)^{2}+e\rho\psi+g-Kv\rho\kappa x\right]\nonumber \\
 &  & +\frac{1}{3}\mathcal{S}_{0}Eh^{3}\kappa^{2},\label{eq:functional}
\end{eqnarray}
\begin{equation}
g=k_{B}T\left[c_{+}\ln(\frac{c_{+}}{c_{\mathrm{max}}})+(c_{\mathrm{max}}-c_{+})\ln(1-\frac{c_{+}}{c_{\mathrm{max}}})\right]\label{eq:entropy_density}
\end{equation}
where $\mathcal{S}$ and $\mathcal{S}_{0}$ are the contact surface area and projected area of the electrodes, respectively, \footnote{For a poor contact between the electrode and the membrane, it is possible that $S<S_{0}$. For a rough electrode surface with good contact, $S>S_{0}$.}; $h$ is the half-separation of electrodes (positioned at $x=-h$ and $x=h$); $\varepsilon$ is the dielectric permittivity (assumed to be constant across the membrane); $\psi(x)$ is the electrostatic mean-field potential distribution; $\rho(x)=c_{+}(x)-c_{0}$ is the local excess number density of cations; $K$ is the bulk modulus of the polymer matrix; $v$ is the excess volume associated with each cation (including solvation shell, if present); $\kappa$ is the curvature of the actuator; $E$ is the elastic modulus of the whole construction including the metallic plates, which in the first approximation can be considered as the volume weighted average of the Young's moduli of the metal and the polymer \cite{Nemat-Nasser:2002zr}; $c_{\mathrm{max}}$ is the maximal local concentration of cations \footnote{Gaussian units are used throughout the article.}.

The first term in Eq. \ref{eq:functional} is the self-energy of the electric field plus that of the battery, keeping a defined potential across the plates. The second term accounts for the electrostatic energy of all ions interacting with each other and with the external field. The third term (see Eq. \ref{eq:entropy_density}) is the entropy of the standard 
\textquotedblleft{}lattice-gas model\textquotedblright{}
\cite{AAK1}, including the volume of ions 
\footnote{In the dilute (Gouy-Chapman, non-interacting) limit $c_{0}<<c_{\mathrm{max}}$, Eq. \ref{eq:entropy_density} reduces to the entropy density of an ideal gas $g_{GC}=k_{B}T\left[c_{+}\ln(c_{+}/c_{0})-c_{+}\right]$.}. The fourth term is the pressure-volume work that provides the coupling between mechanical bending and
the ion concentration profile. Indeed, $Kv\rho$ is the pressure exerted by ions and $\kappa x$ the elongation of the device due to curvature at a distance $x$ away from the centre ($x>0$ in extension; $x<0$ in compression). This effective pressure-volume work characterises ion-induced \emph{swelling}. This term is proportional to the sum of first moments of the charge distributions at each electrode, reflecting the fact that the more polarized the charge distribution, the more favourable bending is and vice versa. The fifth term, outside the integral, is the energetic cost of bending an Euler-Bernoulli beam.\cite{Landau} Note that the beam is assumed to be much longer than its width, precluding Gaussian curvature with only 1-dimensional bending.

All in all, Eqs. \ref{eq:functional},\ref{eq:entropy_density} describe several competing physical effects: the electrostatic and entropic costs of building up ions near one electrode and depleting them at the other. Bending relieves the energy of charge accumulation but the extent of bending is inhibited by the mechanical energy needed to deform the material. 


Minimising the free energy functional (Eqs. \ref{eq:functional},\ref{eq:entropy_density})
with respect to $\rho$, constraining the total number of ions,
gives the electrochemical potential of the mobile ions
$\mu_{+}=e\psi+k_{B}T\ln[c_{+}/(c_{\mathrm{max}}-c_{+})]-Kv\kappa x$. 
By equating $\mu_{+}$ to its value in the bulk, \footnote{As the separation between the electrodes is much greater than the width of the double layers at the electrodes, the bulk of the system is electroneutral and is characterised by constant chemical potential.}
$\mu_{0}=k_{B}T\ln[c_{0}/(c_{\mathrm{max}}-c_{0})]$, 
we obtain the cation distribution (Fermi-like if $\kappa=0$) \cite{AAK1} 
\begin{equation}
c_{+}=\frac{c_{\mathrm{max}}}{\left[\frac{c_{\mathrm{max}}}{c_{0}}-1\right]\exp{\left[\frac{e\psi-Kv\kappa x}{k_{B}T}\right]}+1}. \label{eq:conc_profile}
\end{equation}
Consequently, minimisation of Eqs. \ref{eq:functional} and \ref{eq:entropy_density}
gives a (dimensionless) modified Poisson-Fermi equation
\begin{equation}
\frac{d^{2}y}{dX^{2}}=\frac{\left(1-\gamma\right)\left(e^{y}-1\right)}{(1-\gamma)e^{y}+\gamma},\label{eq:Poisson-Boltzmann}
\end{equation}
with dimensionless variables $X\equiv x/l_{D}$,  $y\equiv e\psi/k_{B}T-p\kappa X\, l_{D}$ where  $p\equiv Kv/k_{B}T$, ion crowding parameter $\gamma\equiv c_{0}/c_{\mathrm{max}}$, Debye length $l_{D}={(4\pi l_{B} c_0)}^{-1/2}$ and Bjerrum length $l_{B}=e^2/\varepsilon k_{B}T$.
Equation \ref{eq:Poisson-Boltzmann} is solved with boundary conditions $y'=y=0$ at $x=0$, corresponding to vanishing electric field and potential in the bulk. These conditions are true for widely separated electrodes, so that the electrical double layer does not overlap the depletion layer, maintaining electroneutrality in the bulk; warranted unless the membrane is nanoscale thin. One can then combine two semi-infinite solutions corresponding to the anode and cathode. Taking the first integral of Eq. \ref{eq:Poisson-Boltzmann}, gives the expression for the electric field: 
\begin{equation}
\left|\frac{\mathrm{d}y}{\mathrm{d}X}\right|=\sqrt{y+\frac{1}{\gamma}\ln\left(\gamma e^{-y}-\gamma+1\right)}.\label{eq:electric_field}
\end{equation}

Minimising Eq. \ref{eq:functional} with respect to $\kappa$ with $H\equiv h/l_{D}$ and $\tilde{\rho}\equiv (c_{+}/c_{0})-1$, gives the dimensionless curvature:
\begin{eqnarray}
\varkappa\equiv\kappa h &=& \frac{3vK}{2Eh^{2}}\frac{S}{S_{0}}l_{D}^{2}c_{0}\int_{-H}^{H}X\tilde{\rho}\,\mathrm{d}X  \nonumber \\
                    &\approx & \frac{3vK}{2Eh}\frac{S}{S_{0}}l_{D}c_{0}\left(\int_{0}^{H}\tilde{\rho}\,\mathrm{d}X-\int_{-H}^{0}\tilde{\rho}\,\mathrm{d}X\right).\label{eq:laplace}
\end{eqnarray}
The approximate equality is justified as the range of the charge density variation is much shorter compared to the separation between the electrodes~\footnote{Equation \ref{eq:laplace} is equivalent to the assumption that bending moment $\sim \int_{-H}^H X {\tilde{\rho}}\; \mathrm{d}X$ \cite{Porfiri:2008hs, Wallmersperger:2007bu}.}. Hence, $\varkappa\thicksim Q$ where $Q$ is the charge per unit cross-section area stored in both the double layer and depletion layer, consistent with experimental results \cite{Akle:2005gh}. The integrals in Eq. \ref{eq:laplace} can then be evaluated using the Gauss law \cite{Landau1}
which relates surface charge densities at each of the electrodes to electric inductions. Since the electrical induction (Eq. \ref{eq:electric_field}) contains itself $\varkappa$ through the definition of $y$, a transcendental
equation for $\varkappa$ emerges:
\begin{eqnarray}
\varkappa & = & \alpha\left\{ \sqrt{2\left[V_{1}-p\varkappa+\frac{1}{\gamma}\ln\left(\gamma e^{-V_{1}+p\varkappa}-\gamma+1\right)\right]}\right.\nonumber \\
 &  & \left.+\sqrt{2\left[-V_{2}+p\varkappa+\frac{1}{\gamma}\ln\left(\gamma e^{V_{2}-p\varkappa}-\gamma+1\right)\right]}\right\} \label{eq:curvature_coltage}
\end{eqnarray}
Here $\alpha$, the dimensionless \emph{electroactuation number}, is the key cumulative parameter for the actuation strength \footnote{The second equality is obtained using the definition of Debye length $l_{D}$, yielding $\alpha \thicksim c_{0}^{1/2}$.}.  
\begin{equation}
\alpha \equiv\frac{3}{2}\frac{K}{E}\frac{l_{D}}{h}\frac{S}{S_0} c_{0} v \equiv \frac{3}{4 \sqrt{\pi}}\frac{K}{E}\frac{S}{S_0}\frac{v}{h}\frac{{c_0}^{1/2}}{{l_{B}}^{1/2}}. \label{eq:gamma} 
\end{equation}
The response vanishes when the membrane is soft or the plates are rigid ($K/E\to0$). 
Here, the dimensionless potential drops $V$, $V_{1}$ and $V_{2}$
are, respectively, the total voltage, and the potential drops near the positive and negative electrodes, all normalized to $k_{B}T/e$, obtained via equating the induction, thus the surface charge density, on both electrodes (this transforms the system into a constant voltage ensemble): 
\begin{equation}
V_{1}=V-V_{2}=p\varkappa+\ln\frac{1-\exp\left[\left(1-\gamma\right)\left(V-2p\varkappa\right)\right]}{\left(1/\gamma-1\right)\left(\exp\left[-\gamma\left(V-2p\varkappa\right)\right]-1\right)}.\label{eq:potential_distribution}
\end{equation}

Expanding Eq. \ref{eq:curvature_coltage} in the low voltage regime gives $V_{1}\approx V_{2}\approx V/2$ and $\varkappa\approx\alpha\sqrt{\left(1-\gamma\right)}V$.
In the opposite large voltage limit, $V_{1}\approx\left(1-\gamma\right)V$
and $\varkappa\approx\alpha\sqrt{8\left(1-\gamma\right)V}$. Figure 3 shows the full numerical solution of Eq. \ref{eq:curvature_coltage}, where these two limits are recovered. This solution is replicated by the interpolation formula  
\begin{equation}
\varkappa=\alpha\sqrt{1-\gamma}\frac{V}{\sqrt{\frac{V}{8}+1}}\label{eq:Intepolation}
\end{equation}
matching the two limits.
As seen from Eq. \ref{eq:Intepolation}, the crossover from linear to non-linear regimes occurs at $V=8$. At 300K, this corresponds to 0.2 Volts (see Fig. 3).
This crossover has been observed experimentally in single-mobile-charge systems \cite{Mariappan:2010qz,Wallmersperger:2008yk} and was rationalized in ref \cite{Colby2011}. 
Rearranging Eq. \ref{eq:Intepolation} yields a simple law for the treatment of experimental data: 
${(V/\varkappa)}^{2}=\alpha^{2}\left(1-\gamma\right)\left(1+\frac{V}{8}\right)$
which gives a straight line plotting ${(V/\varkappa)}^{2}$
vs $V$. Experimental tests of this law and its $1/8$ ratio between the slope and intercept are needed.%

To illustrate how the full solution of Eq. \ref{eq:curvature_coltage} works, consider an example: 
$1\,\mathrm{\mu m}$ thick Nafion of volume fraction $0.6$, with volume fraction of mobile cationic counterions $0.1$ and ethylene carbonate solvent $0.3$ with "typical" physical characteristics (dielectric constant $\varepsilon\thicksim10$  \footnote{The dielectric constant of the system is estimated by the effective medium theory as for a composite mixture of non-polar (polymer) and polar (solvent) phases using the Bruggeman formula \cite{BERGMAN:1978dk}.}, ion excess volume 
$v=0.3\,\mathrm{nm^{3}}$ \footnote{The volume of the ion is taken as the volume of 1-butyl-3-methylimidazolium (BMIm) \cite{Wang:2011ly}, a typical bulky ion used in electroactuators, $v=0.3\,\mathrm{nm^{3}}$.}, ion crowding parameter $\gamma=0.25$ \footnote{The ion crowding parameter $\gamma$ is estimated here from the simple relation $\gamma\sim\frac{\varphi_{ion}}{\varphi_{ion}+\varphi_{solv}}=0.25$
where $\varphi_{solv}$ and $\varphi_{ion}$ are the volume fractions of solvent and charge carrier respectively, as densest packing of ions near the electrode correspond to total displacement of solvent. In typical conditions $0.2\leq\gamma\leq0.3$.} and $K/E=13$ for Nafion \cite{Li:2000rm}).   

\begin{figure}
\includegraphics[scale=0.21]{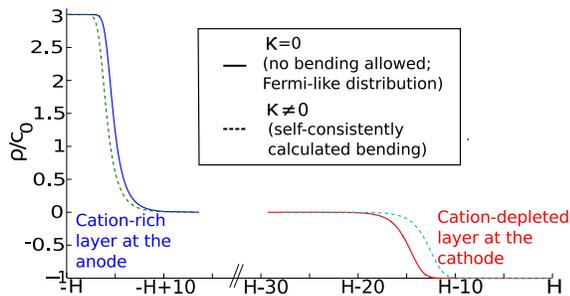}
\caption{Bending reduces the charge separation between the electrodes. Charge density profiles (Eq. \ref{eq:conc_profile}) are plotted as functions of the distance from the corresponding electrode within the proposed self-consistent theory at applied 4 Volts and for $\alpha=1.3\times10^{-3}$, $\gamma=0.25$ and $S/S_{0}=5$. 
$2H\equiv 2hl_{D}^{-1}$, is the membrane thickness measured
in units of Debye length. \label{fig:The-concentration-profile} }
\end{figure}

Figure \ref{fig:The-concentration-profile} shows that the effect of curvature on the concentration profile
is to reduce ion polarization at the electrodes. This is expected, as bending relieves the steric and electrostatic strain induced by the accumulation and depletion of mobile ions near the electrodes. Bending opens extra volume close to the negative electrode, lowering the local concentration of cations. Conversely at the positive electrode, bending decreases the extent of the depletion layer. 
\begin{figure}
\includegraphics[scale=0.21]{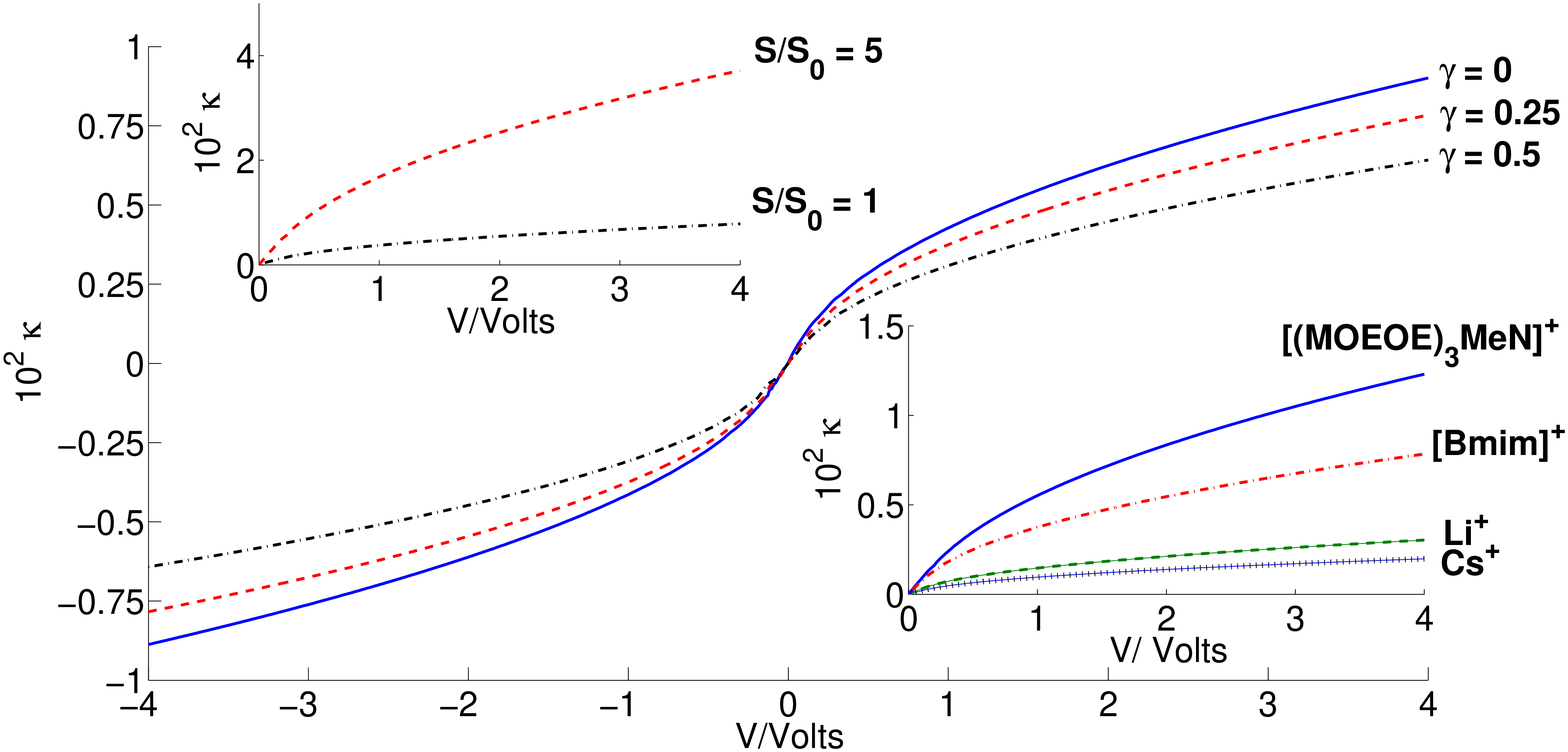}

\caption{Dimensionless curvature $\varkappa$ as a function of applied voltage. The main panel demonstrates the effect of ion crowding parameter $\gamma$ at fixed $\alpha$. The insets show two of the many ways to vary $\alpha$; the effects of roughness factor $S/S_{0}$ and of the effective volume of ion for $\gamma=0.25$, $S/S_{0}=1$ and $v=V_{\mathrm{[BMIm]^{+}}}$.
The volumes of ions used are $V_{\mathrm{[BMIm]^{+}}}=0.3\,\mathrm{nm^{3}}$
\cite{Wang:2011ly}, $V_{\mathrm{[(MOEOE)_{3}MeN]^{+}}}=1\,\mathrm{nm^{3}}$
\cite{Colby2011}, and hydrated ion volumes $V_{\mathrm{Li^{+}}}=0.23\,\mathrm{nm^{3}}$, 
$V_{\mathrm{Cs^{+}}}=0.15\,\mathrm{nm^{3}}$ \cite{Bockris1072}.
Other physical characteristics are the same as Figure 2. }
\end{figure}

Figure 3 demonstrates that larger electrode/membrane contact area and larger volume of the charge carrier enhance bending. More charge is stored for higher surface area electrodes and crowding in the double layer increases with ion size 
\footnote{Compression of the ionic cloud is necessary for the generation of stress and the ion crowding parameter $\gamma$ governs the compressibility thus controls the actuation performance of the system. Indeed at $\gamma=1$, distribution of ions cannot compress and the actuation response is zero.}. 
To achieve a substantial curvature of $\kappa=1 \mathrm{cm}^{-1}$ at an applied voltage of  4V, assuming $\gamma=0.25$, $\alpha/h$ must be greater than $0.03 \mathrm{cm}^{-1}$. Using the above physical characteristics, the separation between the electrodes must be less than $10 \mathrm{\mu m}$. 

This simple model has no hysteresis or dissipation effects; the electroactuator response is fully reversible.
Experimentally, hysteresis may occur due to non-idealities
such as Joule heating, leakage of solvent outside the electrode, irreversible mechanical deformation of the ionic polymer metal composite \cite{Shahinpoor2001,Chen2005} and the current from trace water hydrolysis. Nanoporous electrodes \cite{Kondrat:2011qm,Kondrat:2011bx} are also not considered; surface roughness is assumed to occur on a scale much larger than the Debye length.  

\emph{Dynamics} - The simplest results can be obtained assuming:
(i) the mechanical response time of the polymer is much
faster than the charging of the double layer, (ii) the height of
the electrode roughness is much smaller than the electrode-electrode separation and (iii) the dimensions of all channels are much larger than both the hydrated ion size and the Debye length. In this case, the response dynamics are solely determined by the migration of ions through the bulk and an equilibrium-like charging of the electrical double layers. The theory of double layer charging in single-mobile-charge-carrier systems has been considered in detail \cite{Grahame1947,AAK1}. Applying this to the case of electro-actuation, in response to a dimensionless step potential $U_{0} (\equiv e V_0/k_{B}T)$ imposed at $t=0$, we find an equation for $\varkappa(t)$ \footnote{The equations $V(\varkappa)\approx\varkappa^{2}/\left[8\alpha^{2}(1-\gamma)\right]$
at large voltages and $V(\varkappa)\approx\varkappa/\left[\alpha\sqrt{1-\gamma}\right]$
at low voltages are used.}, 
\begin{equation}
t=\frac{\tau}{\alpha}\int_{0}^{\varkappa(t)}\frac{\mathrm{d}\varkappa}{U_{0}-V(\varkappa)}.
\end{equation}
Here $\tau=R_{0}C_{0}$;  $R_{0}=2h/(S_{0}\sigma)=2hk_{B}T/(S_{0}c_{0}e^{2}D)$ is the resistance of the bulk ($\sigma$ is the ionic conductivity and $D$ is the diffusion coefficient of the ion); 
$C_{0}=\varepsilon S/(8\pi l_{D})$ is the linear Debye capacitance, giving 
\begin{equation}
\tau=\frac{hl_{D}}{D}\frac{S}{S_{0}}.
\end{equation}

Equation 11 is solved analytically in the limits of linear response ($U_0<8$) and strongly nonlinear response:
\begin{equation}
\varkappa(t)= \alpha \sqrt{1-\gamma}
\begin{cases}
U_{0} (1-\exp{\left[\frac{-t}{\tau \sqrt{1-\gamma}}\right]}) \qquad \text{for } U_{0} < 8 \\
\sqrt{8 U_{0} } \mathrm{tanh}\left[\frac{t}{\tau \sqrt{8(1-\gamma)/U_{0}}}\right]  \text{for } U_{0} > 8 . 
\end{cases}
\end{equation}

The two forms coincide at short times, where the initial rate of bending $\alpha U_{0}/\tau$ is independent of $\gamma$, since crowding of ions requires time to develop. The maximum final bending (Eq. \ref{eq:Intepolation}) scales as $\alpha \sqrt{1-\gamma}$ which is maximal at $c_{0}=c_{\mathrm{max}}/2$ (or $\gamma=1/2$). Hence, a vital prediction of our model is that single-ion actuators could be improved by roughly doubling the ion exchange capacity \footnote{Ion exchange capacity refers to the density of ionic groups attached to the polymer (and their counterions) in any single-ion conducting ionomer, see J. A. Kerres, J. Membrane Sci. 185, 3 (2001).} of Nafion ionomers. The relation between excess ion volume and actuation rate can be obtained by noting that in the linear response regime, $\varkappa\sim v t/R_{0}$ with $R_{0}\sim v^{1/3}$ \footnote{The latter is Stoke's Law type scaling. It may break down at the limit of large concentration of ions and non-spherical ions \cite{Vorotyntsev2010}.}. This gives the counter-intuitive prediction $\varkappa\sim v^{2/3}t$ that the actuation becomes more rapid with larger ions!  In the main part of Figure 4, the full $\varkappa(t)$ curve, calculated via Eq. 13, is compared with experimental data  \cite{Leo2007,FarinholtLeo2008}. 

\begin{figure}
\includegraphics[scale=0.23]{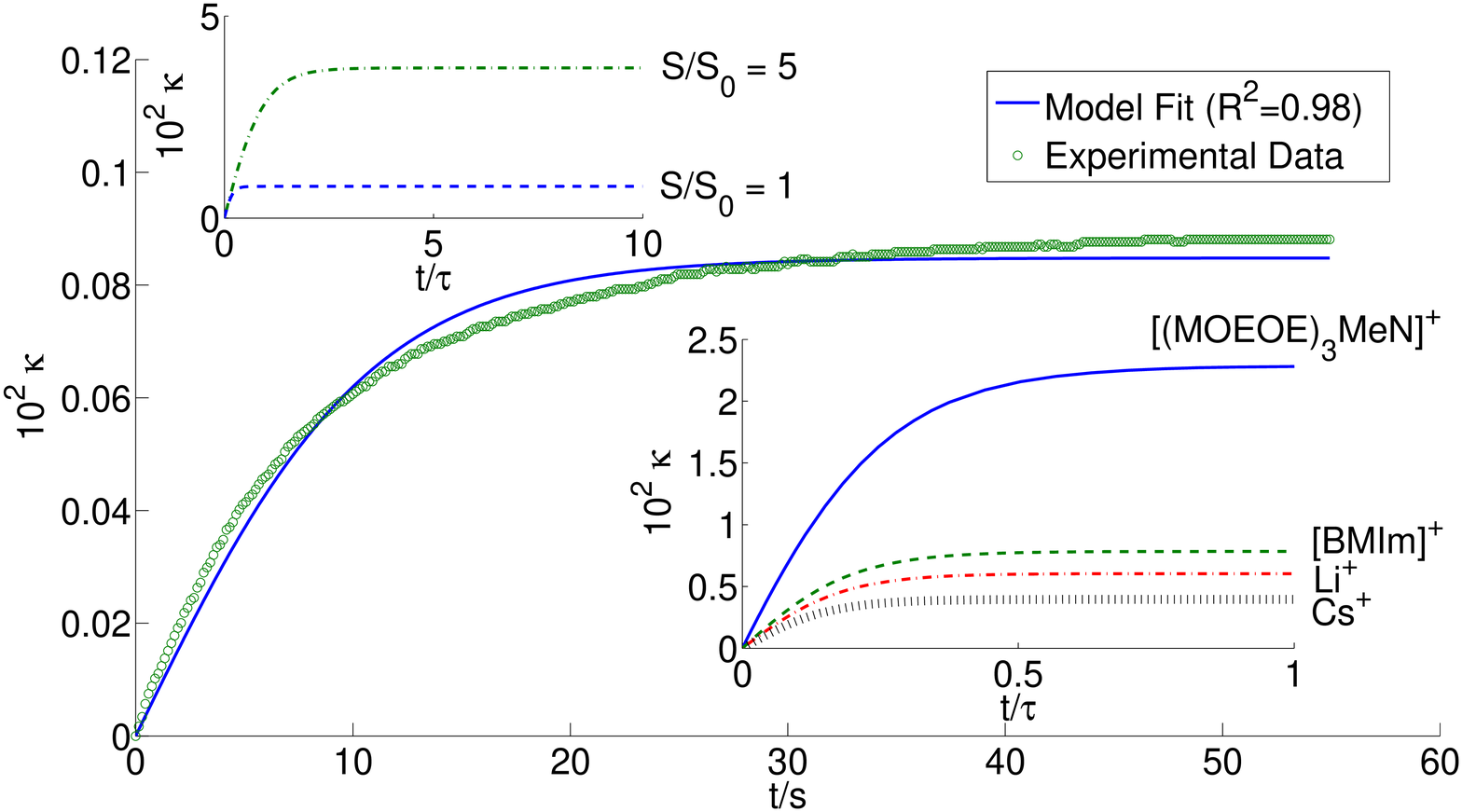}

\caption{The main panel shows fitting of experimental data taken from refs \cite{Leo2007,FarinholtLeo2008} with the non-linear response case of Eq. 13.  There, a rectangular Nafion membrane of dimension $4\times35\times0.2 \mathrm{mm}$
was coated with a platinum electrode and then treated with a thin gold overlayer. The counterion was $\mathrm{Li^{+}}$ with glycerol as solvent. A $1\mathrm{V}$ applied potential ($U_{0}=40$) induced bending. Assuming $\gamma=0.25$, the two-parameter fit to Eq. 13 yields $\alpha=5.6\times10^{-5}$ and $\tau=28s$. The two insets show the effects of the electrode contact surface area and ion size on actuation response, from a $4 \mathrm{V}$ step voltage (see text and the Figure 3 caption for other parameters used). }
\end{figure}

\emph{Conclusion} - A self-consistent theory of single-ion conducting electroactuators is presented, in which the curvature is considered on the same footing as the concentration profiles, modelling both actuation and sensing. The model assumes a strong coupling of ion build-up and depletion with local volume that forces bending. The theory reveals the roles of all physical characteristics but most importantly, they group themselves into three cumulative parameters: (i) dimensionless response coefficient - the \emph{electroactuation number} $\alpha$ (Eq. \ref{eq:gamma}), 
(ii) ion crowding parameter $\gamma$ and 
(iii) the relaxation time $\tau$ for dynamics (Eq. 12). These three parameters have clear constituents, each of which can be independently measured or estimated.  Other results are as follows: 
\newline - At large voltages, both curvature and charge accumulation increase as $\sqrt{V}$, owing to ion crowding.
\newline - For a given voltage, increasing the surface roughness, volume of ion and decreasing the electrode-electrode separation all increase bending. 
\newline - Plotting experimental data as $\left(V/\varkappa\right)^{2}$ vs. $V$ is predicted (Eq. \ref{eq:Intepolation}) to give a straight line with a slope 8 times smaller than the intercept. 
\newline - The dynamic response is described by simple analytical formulae (Eqs. 12 and 13).  
\newline - Electroactuation number $\alpha$ allows a scaling approach for rational design of electroactuators. \footnote{This is applicable in full to the equilibrium response. The effects shown in the insets of Fig. 3 can be accounted for by pure rescaling of $\alpha$. It is trickier in dynamics, however. Electroactuation number $\alpha$ does determine the electroactuation amplitude, but only a subset of the physical characteristics that comprise $\alpha$ enter the characteristic time $\tau$, which also involves the ion diffusion coefficient $D$, see Eq. 12.} 
\begin{acknowledgments}
We thank A. Isaacs, S. Kondrat and Q. Zhang for insightful discussions.
This work is supported in part by the Grant EP/H004319/1 of the Engineering and Physical Science Research Council, Leverhulme Visiting Professorship to RHC, as well as U.S. Army Research Office under Grant No. W911NF-07-1-0452 Ionic Liquids in Electro-Active Devices (ILEAD) MURI.
\end{acknowledgments}
\bibliography{refences_PRL}

\end{document}